# Fokker - Planck equation in curvilinear coordinates

*Igor A. Tanski*
*Moscow, Russia*
povorot2@infoline.su

V&T

*ABSTRACT*

The aim of this paper is to rewrite the Fokker - Planck equation according to transformation of space coordinates. This is nontrivial problem, because transformation of space coordinates induces transformation of velocities. We can use covariant, contravariant or physical velocity components as independent variables in curvilinear coordinate system. These 3 possibilities are considered in this paper and 3 kinds of Fokker - Planck equation in curvilinear coordinates are formulated.

**INTRODUCTION**

Fokker - Planck equation in natural variables is

$$\frac{\partial n}{\partial t} + u_x \frac{\partial n}{\partial x} + u_y \frac{\partial n}{\partial y} + u_z \frac{\partial n}{\partial z} - \alpha \left( \frac{\partial}{\partial u_x}(nu_x) + \frac{\partial}{\partial u_y}(nu_y) + \frac{\partial}{\partial u_z}(nu_z) \right) = k \left( \frac{\partial^2 n}{\partial u_x^2} + \frac{\partial^2 n}{\partial u_y^2} + \frac{\partial^2 n}{\partial u_z^2} \right). \quad (1)$$

where natural variables are defined as

$x, y, z$ - Carthesian coordinates;

$u_x, u_y, u_z$ - velocities.

We shall use index notation.
Index notation for natural variables is:

$$y^1 = x, y^2 = y, y^3 = z.$$

$$u^1 = u_x, u^2 = u_y, u^3 = u_z.$$

Using this notation, we can write the equation (1) in the form

$$\frac{\partial n}{\partial t} + u^i \frac{\partial n}{\partial y^i} - \alpha \, u^i \frac{\partial n}{\partial u^i} - 3\alpha n = k \frac{\partial}{\partial u^i}\left(\frac{\partial n}{\partial u^i}\right). \quad (2)$$

We shall consider natural variables $y_i$, $u_i$ as functions of curvilinear coordinates $x^i$ and corresponding velocities $v^i$. The rule (see [1]) for calculation of velocities components for curvilinear coordinates is

$$v^i = \frac{\partial x^i}{\partial y^k} u^k. \tag{3}$$

where $v^i$ - contravariant components of velocity vector, and

$$v_i = \frac{\partial y^k}{\partial x^i} u^k \tag{4}$$

where $v_i$ - covariant components of velocity vector.

We use usual summation convention: indices that occur twice are considered to be summed over. To prevent summation we underline indices.

The covariant components of metric tensor are by definition

$$g_{mn} = \frac{\partial y^i}{\partial x^m} \frac{\partial y^i}{\partial x^n}. \tag{5}$$

The contravariant components of metric tensor are by definition

$$g^{mn} = \frac{\partial x^m}{\partial y^i} \frac{\partial x^n}{\partial y^i}. \tag{6}$$

Christoffel's symbol is called

$$\Gamma_{p,mn} = \frac{1}{2}\left(\frac{\partial g_{np}}{\partial x^m} + \frac{\partial g_{pm}}{\partial x^n} - \frac{\partial g_{mn}}{\partial x^p}\right) \tag{7}$$

and

$$\Gamma^p_{mn} = g^{pq}\Gamma_{q,mn}. \tag{8}$$

Using these definitions, we can write second derivatives of curvilinear coordinates in the following form:

$$\frac{\partial^2 x^r}{\partial y^k \partial y^l} = -\frac{\partial x^s}{\partial y^k}\frac{\partial x^t}{\partial y^l}\Gamma^r_{st}, \tag{9}$$

and second derivatives of Cartesian coordinates - in the form

$$\frac{\partial^2 y^i}{\partial x^m \partial x^n} = \frac{\partial y^i}{\partial x^q}\Gamma^q_{mn}. \tag{10}$$

In curvilinear coordinates we can choose between covariant and contravariant velocity components as independent variables. There is the third possibility for orthogonal coordinates - physical components of velocities.

## 1. Contravariant components of velocities

Expanding expression for the $n$ derivatives, we have

$$\frac{\partial n}{\partial y^i} = \frac{\partial n}{\partial x^k}\frac{\partial x^k}{\partial y^i} + \frac{\partial n}{\partial v^k}\frac{\partial v^k}{\partial y^i}. \tag{11}$$

$$\frac{\partial n}{\partial u^i} = \frac{\partial n}{\partial x^k}\frac{\partial x^k}{\partial u^i} + \frac{\partial n}{\partial v^k}\frac{\partial v^k}{\partial u^i}. \tag{12}$$

if we choose contravariant velocity components as independent variables.

$$\frac{\partial v^k}{\partial y^i} = \frac{\partial}{\partial y^i}\left(u^m \frac{\partial x^k}{\partial y^m}\right) = u^m \frac{\partial^2 x^k}{\partial y^m \partial y^i} = -u^m \frac{\partial x^s}{\partial y^m}\frac{\partial x^t}{\partial y^i}\Gamma^k_{st}; \tag{13}$$

$$\frac{\partial v^k}{\partial u^i} = \frac{\partial}{\partial u^i}\left(u^m \frac{\partial x^k}{\partial y^m}\right) = \frac{\partial x^k}{\partial y^i}; \tag{14}$$

because $\frac{\partial x^k}{\partial u^i} = 0$.

.PP Hence, we have following expressions for $n$ derivatives

$$\frac{\partial n}{\partial y^i} = \frac{\partial n}{\partial x^k}\frac{\partial x^k}{\partial y^i} - \frac{\partial n}{\partial v^k} u^m \frac{\partial x^s}{\partial y^m}\frac{\partial x^t}{\partial y^i}\Gamma^k_{st} = \frac{\partial x^k}{\partial y^i}\left(\frac{\partial n}{\partial x^k} - \frac{\partial n}{\partial v^t}u^m \frac{\partial x^s}{\partial y^m}\Gamma^t_{sk}\right), \tag{15}$$

$$\frac{\partial n}{\partial u^i} = \frac{\partial n}{\partial v^k}\frac{\partial x^k}{\partial y^i}, \tag{16}$$

if we choose contravariant velocity components as primary variables.

Substituting (15-16) for &n& derivatives in equation (2), we obtain

$$\frac{\partial n}{\partial t} + u^i \frac{\partial x^k}{\partial y^i}\left(\frac{\partial n}{\partial x^k} - \frac{\partial n}{\partial v^t}u^m\frac{\partial x^s}{\partial y^m}\Gamma^t_{sk}\right) - \alpha\, u^i \frac{\partial n}{\partial v^k}\frac{\partial x^k}{\partial y^i} - 3\,\alpha\, n = k\, \frac{\partial x^l}{\partial y^i}\frac{\partial}{\partial v^l}\left(\frac{\partial x^k}{\partial y^i}\frac{\partial n}{\partial v^k}\right). \tag{17}$$

$$\frac{\partial n}{\partial t} + v^k \frac{\partial n}{\partial x^k} - v^k \frac{\partial n}{\partial v^t} v^s \Gamma^t_{sk} - \alpha\, v^k \frac{\partial n}{\partial v^k} - 3\,\alpha\, n = k\, \frac{\partial x^l}{\partial y^i}\frac{\partial x^k}{\partial y^i}\frac{\partial^2 n}{\partial v^l \partial v^k}. \tag{18}$$

Finally, equation (2) in curvilinear coordinates and with contravariant velocities as independent variables, is

$$\frac{\partial n}{\partial t} + v^k \frac{\partial n}{\partial x^k} - \Gamma^k_{pq} v^p v^q \frac{\partial n}{\partial v^k} - \alpha\, v^k \frac{\partial n}{\partial v^k} - 3\,\alpha\, n = k\, g^{lk} \frac{\partial^2 n}{\partial v^l \partial v^k}. \tag{19}$$

## 2. Covariant components of velocities

Expanding expression for the $n$ derivatives, we have

$$\frac{\partial n}{\partial y^i} = \frac{\partial n}{\partial x^k}\frac{\partial x^k}{\partial y^i} + \frac{\partial n}{\partial v_k}\frac{\partial v_k}{\partial y^i}. \tag{20}$$

$$\frac{\partial n}{\partial u^i} = \frac{\partial n}{\partial x^k}\frac{\partial x^k}{\partial u^i} + \frac{\partial n}{\partial v_k}\frac{\partial v_k}{\partial u^i}. \tag{21}$$

if we choose covariant velocity components as primary variables.

$$\frac{\partial v_k}{\partial y^i} = \frac{\partial}{\partial y^i}\left(u^m \frac{\partial y^m}{\partial x^k}\right) = u^m \frac{\partial^2 y^m}{\partial x^k \partial x^l} \frac{\partial x^l}{\partial y^i} = u^m \frac{\partial y^m}{\partial x^q} \Gamma^q_{kl} \frac{\partial x^l}{\partial y^i} = v_q \Gamma^q_{kl} \frac{\partial x^l}{\partial y^i} ; \qquad (22)$$

$$\frac{\partial v_k}{\partial u^i} = \frac{\partial}{\partial u^i}\left(u^m \frac{\partial y^m}{\partial x^k}\right) = \frac{\partial y^i}{\partial x^k} . \qquad (23)$$

because $\dfrac{\partial x^k}{\partial u^i} = 0$.

Hence, we have the following expressions for *n* derivatives

$$\frac{\partial n}{\partial y^i} = \frac{\partial n}{\partial x^k} \frac{\partial x^k}{\partial y^i} + \frac{\partial n}{\partial v_k} v_q \Gamma^q_{kl} \frac{\partial x^l}{\partial y^i} = \frac{\partial x^k}{\partial y^i}\left(\frac{\partial n}{\partial x^k} + \frac{\partial n}{\partial v_l} v_q \Gamma^q_{lk}\right), \qquad (24)$$

$$\frac{\partial n}{\partial u^i} = \frac{\partial n}{\partial v_k} \frac{\partial y^i}{\partial x^k} \qquad (25)$$

if we choose covariant velocity components as primary variables.

Substituting (24-25) for *n* derivatives in equation (2), we obtain

$$\frac{\partial n}{\partial t} + u^i \frac{\partial x^k}{\partial y^i}\left(\frac{\partial n}{\partial x^k} + \frac{\partial n}{\partial v_l} v_q \Gamma^q_{lk}\right) - \alpha\, u^i \frac{\partial n}{\partial v_k} \frac{\partial y^i}{\partial x^k} - 3\,\alpha\, n = k\, \frac{\partial y^i}{\partial x^l} \frac{\partial}{\partial v_l}\left(\frac{\partial y^i}{\partial x^k} \frac{\partial n}{\partial v_k}\right). \qquad (26)$$

$$\frac{\partial n}{\partial t} + v_s g^{sk} \frac{\partial n}{\partial x^k} + v_s g^{sk} \frac{\partial n}{\partial v_l} v_q \Gamma^q_{lk} - \alpha\, v_k \frac{\partial n}{\partial v_k} - 3\,\alpha\, n = k\, \frac{\partial y^i}{\partial x^l} \frac{\partial y^i}{\partial x^k} \frac{\partial^2 n}{\partial v_l \partial v_k} . \qquad (27)$$

Finally, equation (2) in curvilinear coordinates and with covariant velocities as independent variables, is

$$\frac{\partial n}{\partial t} + g^{mk} v_m \frac{\partial n}{\partial x^k} + \Gamma^q_{kl} g^{pl} v_p v_q \frac{\partial n}{\partial v_k} - \alpha\, v_k \frac{\partial n}{\partial v_k} - 3\,\alpha\, n = k\, g_{lk} \frac{\partial^2 n}{\partial v_l \partial v_k} . \qquad (28)$$

## 3. Orthogonal coordinates

In orthogonal coordinates components of metric tensor have the following properties:

$$g_{mn} = g^{mn} = 0; \quad (m \neq n); \qquad (29)$$

$$g_{\underline{ii}} = (H_i)^2; \qquad (30)$$

$$g^{\underline{ii}} = \frac{1}{(H_i)^2} ; \qquad (31)$$

where $H^i$ - Lame coefficients.

Here and after we do not sum on underlined symbols (e.g. $\underline{i}$).

The expressions for Christoffel symbols in orthogonal coordinates are

$$\Gamma_{k,ij} = 0; \quad (i \neq j \neq k). \tag{32}$$

$$\Gamma_{\underline{i},\underline{i}j} = H_{\underline{i}} \frac{\partial H_{\underline{i}}}{\partial x^j}; \quad (i \neq j). \tag{33}$$

$$\Gamma_{j,\underline{ii}} = -H_{\underline{i}} \frac{\partial H_{\underline{i}}}{\partial x^j}; \quad (i \neq j). \tag{34}$$

$$\Gamma_{\underline{i},\underline{ii}} = H_{\underline{i}} \frac{\partial H_{\underline{i}}}{\partial x^{\underline{i}}}. \tag{35}$$

$$\Gamma^k_{ij} = 0; \quad (i \neq j \neq k). \tag{36}$$

$$\Gamma^{\underline{i}}_{\underline{i}j} = \frac{1}{H_{\underline{i}}} \frac{\partial H_{\underline{i}}}{\partial x^j}; \quad (i \neq j). \tag{37}$$

$$\Gamma^j_{\underline{ii}} = -\frac{H_{\underline{i}}}{(H_j)^2} \frac{\partial H_{\underline{i}}}{\partial x^{\underline{i}}}; \quad (i \neq j). \tag{38}$$

$$\Gamma^{\underline{i}}_{\underline{ii}} = \frac{1}{H_{\underline{i}}} \frac{\partial H_{\underline{i}}}{\partial x^{\underline{i}}}. \tag{39}$$

Let us denote by $w^i$ physical components of velocity vector

$$w^{\underline{i}} = H_{\underline{i}} v^{\underline{i}} = \frac{v_{\underline{i}}}{H_{\underline{i}}}. \tag{40}$$

$$w^{\underline{i}} = H_{\underline{i}} \frac{\partial x^{\underline{i}}}{\partial y^k} u^k = \frac{1}{H_{\underline{i}}} \frac{\partial y^k}{\partial x^{\underline{i}}} u^k. \tag{41}$$

Expanding expression for the $n$ derivatives, we have

$$\frac{\partial n}{\partial y^i} = \frac{\partial n}{\partial x^k} \frac{\partial x^k}{\partial y^i} + \frac{\partial n}{\partial w^k} \frac{\partial w^k}{\partial y^i}. \tag{42}$$

$$\frac{\partial n}{\partial u^i} = \frac{\partial n}{\partial x^k} \frac{\partial x^k}{\partial u^i} + \frac{\partial n}{\partial w^k} \frac{\partial w^k}{\partial u^i}. \tag{43}$$

if we choose physical velocity components as independent variables.

$$\frac{\partial w^k}{\partial y^i} = \frac{\partial}{\partial y^i}\left(H_{\underline{k}} \frac{\partial x^{\underline{k}}}{\partial y^m} u^m\right) = u^m\left(H_{\underline{k}} \frac{\partial^2 x^{\underline{k}}}{\partial y^m \partial y^i} + \frac{\partial H_{\underline{k}}}{\partial y^i} \frac{\partial x^{\underline{k}}}{\partial y^m}\right); \tag{44}$$

$$\frac{\partial w^k}{\partial u^i} = \frac{\partial}{\partial u^i}\left(H_{\underline{k}} \frac{\partial x^{\underline{k}}}{\partial y^m} u^m\right) = H_{\underline{k}} \frac{\partial x^{\underline{k}}}{\partial y^i}; \tag{45}$$

because $\dfrac{\partial x^k}{\partial u^i} = 0$.

Let us simplify the expression in parenthesis in (44)

$$H_{\underline{k}} \frac{\partial^2 x^{\underline{k}}}{\partial y^m \partial y^i} + \frac{\partial H_{\underline{k}}}{\partial y^i} \frac{\partial x^{\underline{k}}}{\partial y^m} = -H_{\underline{k}} \frac{\partial x^s}{\partial y^m} \frac{\partial x^t}{\partial y^i} \Gamma^k_{st} + \frac{\partial H_{\underline{k}}}{\partial x^p} \frac{\partial x^p}{\partial y^i} \frac{\partial x^{\underline{k}}}{\partial y^m}. \tag{46}$$

$$H_{\underline{k}} \frac{\partial^2 x^{\underline{k}}}{\partial y^m \partial y^i} + \frac{\partial H_{\underline{k}}}{\partial y^i} \frac{\partial x^{\underline{k}}}{\partial y^m} = -H_{\underline{k}} \frac{\partial x^s}{\partial y^m} \frac{\partial x^t}{\partial y^i} \Gamma^k_{st} + H_{\underline{k}} \Gamma^k_{\underline{k}p} \frac{\partial x^p}{\partial y^i} \frac{\partial x^{\underline{k}}}{\partial y^m}. \tag{47}$$

$$H_{\underline{k}} \frac{\partial^2 x^{\underline{k}}}{\partial y^m \partial y^i} + \frac{\partial H_{\underline{k}}}{\partial y^i} \frac{\partial x^{\underline{k}}}{\partial y^m} = -H_{\underline{k}} \frac{\partial x^p}{\partial y^i}\left(\frac{\partial x^s}{\partial y^m} \Gamma^k_{sp} - \Gamma^k_{\underline{k}p} \frac{\partial x^{\underline{k}}}{\partial y^m}\right). \tag{48}$$

If $s = k$ we can cancel first two terms in parenthesis in (48)

$$H_{\underline{k}} \frac{\partial^2 x^{\underline{k}}}{\partial y^m \partial y^i} + \frac{\partial H_{\underline{k}}}{\partial y^i} \frac{\partial x^{\underline{k}}}{\partial y^m} = -H_{\underline{k}} \frac{\partial x^p}{\partial y^i} \frac{\partial x^s}{\partial y^m} \Gamma^k_{sp}. \quad (s \neq k) \tag{49}$$

The situation is not quite usual, so small comment is needed. (49) means, that index $k$ as usual takes values 1, 2, 3. Index $s$ for each given $k$ takes values from the set 1, 2, 3 - except $k$. We sum values of the expression, calculated for these values of indices $k$ and $s$.

Proceeding with our calculations, we note, that in (49) must be $p = k$ or $p = s$ - else $\Gamma^k_{sp} = 0$. So we have

$$H_{\underline{k}} \frac{\partial^2 x^{\underline{k}}}{\partial y^m \partial y^i} + \frac{\partial H_{\underline{k}}}{\partial y^i} \frac{\partial x^{\underline{k}}}{\partial y^m} = -H_{\underline{k}} \frac{\partial x^{\underline{k}}}{\partial y^i} \frac{\partial x^s}{\partial y^m} \Gamma^k_{s\underline{k}} - H_{\underline{k}} \frac{\partial x^s}{\partial y^i} \frac{\partial x^s}{\partial y^m} \Gamma^k_{ss} = \tag{50}$$

$$= -H_{\underline{k}} \frac{\partial x^{\underline{k}}}{\partial y^i} \frac{\partial x^s}{\partial y^m} \frac{1}{H_{\underline{k}}} \frac{\partial H_{\underline{k}}}{\partial x^s} + H_{\underline{k}} \frac{\partial x^s}{\partial y^i} \frac{\partial x^s}{\partial y^m} \frac{H_s}{(H_{\underline{k}})^2} \frac{\partial H_s}{\partial x^{\underline{k}}} =$$

$$= \frac{\partial x^s}{\partial y^m}\left(\frac{\partial x^s}{\partial y^i} \frac{H_s}{H_{\underline{k}}} \frac{\partial H_s}{\partial x^{\underline{k}}} - \frac{\partial x^{\underline{k}}}{\partial y^i} \frac{\partial H_{\underline{k}}}{\partial x^s}\right) \quad (s \neq k)$$

and this is the desired form of expression in parenthesis in (44).

So the expression for derivatives of velocities physical components has the form

$$\frac{\partial w^k}{\partial y^i} = \frac{w^s}{H_s}\left(\frac{\partial x^s}{\partial y^i}\frac{H_s}{H_{\underline{k}}}\frac{\partial H_s}{\partial x^{\underline{k}}} - \frac{\partial x^{\underline{k}}}{\partial y^i}\frac{\partial H_{\underline{k}}}{\partial x^s}\right) \quad (s \neq k) \tag{51}$$

Combining (51) with (42-43), we get:

$$\frac{\partial n}{\partial y^i} = \frac{\partial n}{\partial x^k}\frac{\partial x^k}{\partial y^i} + \frac{\partial n}{\partial w^k}\frac{w^s}{H_s}\left(\frac{\partial x^s}{\partial y^i}\frac{H_s}{H_k}\frac{\partial H_s}{\partial x^k} - \frac{\partial x^k}{\partial y^i}\frac{\partial H_k}{\partial x^s}\right) \quad (s \neq k) \tag{52}$$

$$\frac{\partial n}{\partial u^i} = H_k \frac{\partial n}{\partial w^k}\frac{\partial x^k}{\partial y^i}. \tag{53}$$

These are the expressions for *n* derivatives if we choose physical velocity components as independent variables.

Substituting (52-53) for &n& derivatives in equation (2), we obtain $(s \neq k)$

$$\frac{\partial n}{\partial t} + u^i \frac{\partial x^k}{\partial y^i}\frac{\partial n}{\partial x^k} + u^i \frac{\partial n}{\partial w^k}\frac{w^s}{H_s}\left(\frac{\partial x^s}{\partial y^i}\frac{H_s}{H_k}\frac{\partial H_s}{\partial x^k} - \frac{\partial x^k}{\partial y^i}\frac{\partial H_k}{\partial x^s}\right) - \alpha\, u^i H_k \frac{\partial n}{\partial w^k}\frac{\partial x^k}{\partial y^i} - 3\,\alpha\,n = k\, H_l\,\frac{\partial x^l}{\partial y^i}\frac{\partial}{\partial w^l}\left(H_k \frac{\partial x^k}{\partial y^i}\frac{\partial n}{\partial w^k}\right). \tag{54}$$

Finally, equation (2) in curvilinear coordinates and with physical velocities as independent variables, is

$$\frac{\partial n}{\partial t} + \frac{w^k}{H_k}\frac{\partial n}{\partial x^k} + \frac{\partial n}{\partial w^k}\frac{w^s}{H_s H_k}\left(w^s \frac{\partial H_s}{\partial x^k} - w^k \frac{\partial H_k}{\partial x^s}\right) - \alpha\, w^k \frac{\partial n}{\partial w^k} - 3\,\alpha\,n = k\,\frac{\partial^2 n}{\partial w^i \partial w^i}. \quad (s \neq k) \tag{55}$$

### 4. Cross checking of equations

Equations (19), (28), (55) were obtained from the single source - Fokker - Planck equation (1) for Carthesian coordinates. It is interesting to show, that each of them can be obtained from the others.

To get one equation from another we perform linear substitution on velocities, space variables remain without changes. Derivatives of *n* on velocities are subject to corresponding linear substitution. It is easy to show that terms with derivatives of *n* on velocities in equations (19), (28), (55) are consistent.

For example, let us check the correspondence of terms with the second derivatives on velocities in (19) and (28). We are to prove that

$$g^{lk}\frac{\partial^2 n}{\partial v^l \partial v^k} = g_{lk}\frac{\partial^2 n}{\partial v_l \partial v_k}. \tag{56}$$

But

$$\frac{\partial n}{\partial v^l} = \frac{\partial n}{\partial v_k}\frac{\partial v_k}{\partial v^l} = \frac{\partial n}{\partial v_k} g_{kl}; \tag{57}$$

therefore

$$g^{lk}\frac{\partial^2 n}{\partial v^l \partial v^k} = g^{lk}\frac{\partial^2 n}{\partial v_m \partial v_n} g_{ml} g_{nl} = g_{mn}\frac{\partial^2 n}{\partial v_m \partial v_n} \tag{58}$$

this ends the proof.

It is equally easy to check consistency of terms $v^k \frac{\partial n}{\partial v^k}$.

It is more interesting to show the identity of terms, produced by $u^i \frac{\partial n}{\partial y^i}$ term in (1). For the brevity sake we drop contravariant velocity multiplier before parenthesis and check the equality (compare with (17) and (26))

$$\left(\frac{\partial n}{\partial x^k}\right)_{v^\alpha} - \frac{\partial n}{\partial v^t} v^s \Gamma^t_{sk} = \left(\frac{\partial n}{\partial x^k}\right)_{v_\alpha} + \frac{\partial n}{\partial v_l} v_q \Gamma^q_{lk}. \tag{59}$$

But

$$\left(\frac{\partial n}{\partial x^k}\right)_{v^\alpha} = \left(\frac{\partial n}{\partial x^k}\right)_{v_\alpha} + \frac{\partial n}{\partial v_p} \frac{\partial g_{pm}}{\partial x^k} v^m. \tag{60}$$

Combine (60) with (59) and get

$$\frac{\partial n}{\partial v_p} \frac{\partial g_{pm}}{\partial x^k} v^m - \frac{\partial n}{\partial v^p} g_{pt} v^s \Gamma^t_{sk} = \frac{\partial n}{\partial v_l} v^q \Gamma_{q,kl}, \tag{61}$$

or

$$\frac{\partial n}{\partial v_p} v^m \frac{\partial g_{pm}}{\partial x^k} = \frac{\partial n}{\partial v_p} v^m \Gamma_{p,mk} + \frac{\partial n}{\partial v_p} v^m \Gamma_{m,pk}. \tag{62}$$

Identity $\frac{\partial g_{pm}}{\partial x^k} = \Gamma_{p,mk} + \Gamma_{m,pk}$ follows directly from definition of Christoffel's symbols (7).

Let us check correspondence of $u^i \frac{\partial n}{\partial y^i}$ terms in (19) and (55). We check the equality (compare with (17) and (54))

$$v^k \left(\left(\frac{\partial n}{\partial x^k}\right)_{v^\alpha} - \frac{\partial n}{\partial v^t} v^p \Gamma^t_{pk}\right) = \frac{w^k}{H_k} \left(\frac{\partial n}{\partial x^k}\right)_{w^\alpha} + \frac{\partial n}{\partial w^k} \frac{w^s}{H_s H_k} \left(w^s \frac{\partial H_s}{\partial x^k} - w^k \frac{\partial H_k}{\partial x^s}\right). (s \neq k) \tag{63}$$

But

$$\left(\frac{\partial n}{\partial x^k}\right)_{v^\alpha} = \left(\frac{\partial n}{\partial x^k}\right)_{w^\alpha} + \frac{\partial n}{\partial w^p} \frac{\partial H_p}{\partial x^k} v^p. \tag{64}$$

Combine (64) with (63) and get

$$v^k \left(\frac{\partial n}{\partial w^p} \frac{\partial H_p}{\partial x^k} v^p - H_t \frac{\partial n}{\partial w^t} v^p \Gamma^t_{pk}\right) = \frac{\partial n}{\partial w^k} \frac{w^s}{H_s H_k} \left(w^s \frac{\partial H_s}{\partial x^k} - w^k \frac{\partial H_k}{\partial x^s}\right). (s \neq k) \tag{65}$$

or

$$\frac{\partial n}{\partial w^p} v^k \left( \frac{\partial H_p}{\partial x^k} v^p - H_p v^t \Gamma^p_{tk} \right) = \frac{\partial n}{\partial w^k} \frac{v^s}{H_k} \left( H_s v^s \frac{\partial H_s}{\partial x^k} - H_k v^k \frac{\partial H_k}{\partial x^s} \right), (s \neq k) \quad (66)$$

We drop multiplier $\frac{\partial n}{\partial w^p}$ - this makes index $p$ in the rest of the expression underlined.

$$v^k \left( \frac{\partial H_{\underline{p}}}{\partial x^k} v^{\underline{p}} - H_{\underline{p}} v^t \Gamma^{\underline{p}}_{tk} \right) = \frac{v^s}{H_{\underline{p}}} \left( H_s v^s \frac{\partial H_s}{\partial x^{\underline{p}}} - H_{\underline{p}} v^{\underline{p}} \frac{\partial H_{\underline{p}}}{\partial x^s} \right), (s \neq p) \quad (67)$$

Let us consider index $t$ in the second term of the left part. When it is equal to $p$ we can cancel two first terms, so we have

$$-v^k H_{\underline{p}} v^t \Gamma^{\underline{p}}_{tk} = \frac{v^s}{H_{\underline{p}}} \left( H_s v^s \frac{\partial H_s}{\partial x^{\underline{p}}} - H_{\underline{p}} v^{\underline{p}} \frac{\partial H_{\underline{p}}}{\partial x^s} \right), (s \neq p) \text{ and } (t \neq p) \quad (68)$$

In this case must be $k = t$ or $k = p$ - else $\Gamma^p_{tk}$ is zero

$$-v^t H_{\underline{p}} v^t \Gamma^{\underline{p}}_{tt} - v^{\underline{p}} H_{\underline{p}} v^t \Gamma^{\underline{p}}_{t\underline{p}} = \frac{v^s}{H_{\underline{p}}} \left( H_s v^s \frac{\partial H_s}{\partial x^{\underline{p}}} - H_{\underline{p}} v^{\underline{p}} \frac{\partial H_{\underline{p}}}{\partial x^s} \right), (s \neq p) \text{ and } (t \neq p) \quad (69)$$

This follows directly from the definition of Christoffel's symbols (37-38).

This proves, that equations (19), (28), (55) follows one from another,. they need not to be the sequence of one equation (1) for the Carthesian coordinates. They are valid for curved manifold, where neither Cartesian coordinate system exists no equation (1). So they can be considered as generalization of Fokker - Planck equation (1). The manifold must be Riemannian, because Christoffel's symbols must satisfy (7).

## 5. Spherical coordinates

To give an example, we perform calculations for the case of spherical coordinates. Properties of spherical coordinates are as follows:

$$x = r \sin(\theta) \cos(\phi); \quad y = r \sin(\theta) \sin(\phi); \quad z = r \cos(\theta). \quad (70)$$

$$x^1 = r; \quad x^2 = \theta; \quad x^3 = \phi. \quad (71)$$

$$g_{11} = 1; \quad g_{22} = r^2; \quad g_{33} = r^2 \sin^2(\theta). \quad (72)$$

$$g_{12} = g_{23} = g_{31} = g_{21} = g_{32} = g_{13} = 0. \quad (73)$$

$$g^{11} = 1; \quad g^{22} = \frac{1}{r^2}; \quad g^{33} = \frac{1}{r^2 \sin^2(\theta)}. \tag{74}$$

$$g^{12} = g^{23} = g^{31} = g^{21} = g^{32} = g^{13} = 0. \tag{75}$$

$$H_1 = 1; \quad H_2 = r; \quad H_3 = r \sin(\theta). \tag{76}$$

$$\Gamma_{1,22} = -r; \quad \Gamma_{2,12} = \Gamma_{2,21} = r; \tag{77}$$

$$\Gamma_{1,33} = -r \sin^2(\theta); \quad \Gamma_{3,13} = \Gamma_{3,31} = r \sin^2(\theta); \tag{78}$$

$$\Gamma_{2,33} = -r^2 \sin(\theta) \cos(\theta); \quad \Gamma_{3,23} = \Gamma_{3,32} = r^2 \sin(\theta) \cos(\theta). \tag{79}$$

$$\Gamma^1_{22} = -r; \quad \Gamma^2_{12} = \Gamma^2_{21} = \frac{1}{r}; \tag{80}$$

$$\Gamma^1_{33} = -r \sin^2(\theta); \quad \Gamma^3_{13} = \Gamma^3_{31} = \frac{1}{r}; \tag{81}$$

$$\Gamma^2_{33} = -\sin(\theta) \cos(\theta); \quad \Gamma^3_{23} = \Gamma^3_{32} = \frac{\cos(\theta)}{\sin(\theta)}. \tag{82}$$

Substituting (70-82) in (19), (28) and (55), we obtain 3 kinds of Fokker - Planck equation for the case of spherical coordinates.

The equation (19) in spherical coordinates has the form:

$$\frac{\partial n}{\partial t} + v^1 \frac{\partial n}{\partial r} + v^2 \frac{\partial n}{\partial \theta} + v^3 \frac{\partial n}{\partial \phi} + \tag{83}$$

$$+ r \left( v^2 v^2 + \sin^2(\theta) v^3 v^3 \right) \frac{\partial n}{\partial v^1} + \left( \sin(\theta) \cos(\theta) v^3 v^3 - \frac{2}{r} v^1 v^2 \right) \frac{\partial n}{\partial v^2} - 2 \left( v^1 v^3 + \frac{\cos(\theta)}{\sin(\theta)} v^2 v^3 \right) \frac{\partial n}{\partial v^3} -$$

$$- \alpha \left( v^1 \frac{\partial n}{\partial v^1} + v^2 \frac{\partial n}{\partial v^2} + v^3 \frac{\partial n}{\partial v^3} \right) - 3 \alpha n = k \left( \frac{\partial^2 n}{\partial v^1 \partial v^1} + \frac{1}{r^2} \frac{\partial^2 n}{\partial v^2 \partial v^2} + \frac{1}{r^2 \sin^2(\theta)} \frac{\partial^2 n}{\partial v^3 \partial v^3} \right).$$

The equation (28) in spherical coordinates has the form:

$$\frac{\partial n}{\partial t} + v_1 \frac{\partial n}{\partial r} + \frac{v_2}{r^2} \frac{\partial n}{\partial \theta} + \frac{v_3}{r^2 \sin^2(\theta)} \frac{\partial n}{\partial \phi} - \tag{84}$$

$$+ \frac{1}{r^3} \left( v_2 v_2 + \frac{v_3 v_3}{\sin^2(\theta)} \right) \frac{\partial n}{\partial v_1} + \frac{\cos(\theta) v_3 v_3}{r^2 \sin^3(\theta)} \frac{\partial n}{\partial v_2} -$$

$$- \alpha \left( v_1 \frac{\partial n}{\partial v_1} + v_2 \frac{\partial n}{\partial v_2} + v_3 \frac{\partial n}{\partial v_3} \right) - 3 \alpha n = k \left( \frac{\partial^2 n}{\partial v_1 \partial v_1} + r^2 \frac{\partial^2 n}{\partial v_2 \partial v_2} + r^2 \sin^2(\theta) \frac{\partial^2 n}{\partial v_3 \partial v_3} \right).$$

The equation (55) in spherical coordinates has the form:

$$\frac{\partial n}{\partial t} + w^1 \frac{\partial n}{\partial r} + \frac{w^2}{r} \frac{\partial n}{\partial \theta} + \frac{w^3}{r \sin(\theta)} \frac{\partial n}{\partial \phi} + \quad (85)$$

$$+ \frac{1}{r}\left(w^2 w^2 + w^3 w^3\right)\frac{\partial n}{\partial w^1} + \frac{1}{r}\left(\frac{\cos(\theta)}{\sin(\theta)} w^3 w^3 - w^1 w^2\right)\frac{\partial n}{\partial w^2} - \frac{1}{r}\left(w^1 w^3 + \frac{\cos(\theta)}{\sin(\theta)} w^2 w^3\right)\frac{\partial n}{\partial w^3} -$$

$$- \alpha \left(w^1 \frac{\partial n}{\partial w^1} + w^2 \frac{\partial n}{\partial w^2} + w^3 \frac{\partial n}{\partial w^3}\right) - 3\,\alpha\,n = k\left(\frac{\partial^2 n}{\partial w^1 \partial w^1} + \frac{\partial^2 n}{\partial w^2 \partial w^2} + \frac{\partial^2 n}{\partial w^3 \partial w^3}\right)$$

**DISCUSSION**

Present work is of rather technical kind. Its main result are three kinds of Fokker - Planck equations in curvilinear coordinates (19), (28), (55).

Some conclusions:

1. We emphasize that (19), (28), (55) are valid not only as simple transcription of equation (1) for curvilinear coordinates. They are also valid for curved Riemannian manifold - where no Cartesian coordinates system exist. So they can be considered as generalization of simple Fokker - Planck equation.

2. We see, that in curvilinear coordinates the left side of the equation is quadratic form of velocities (in Cartesian coordinates it was linear form). This gives the possibility to perform the Fourier transform. Fourier transform of the equation is linear PDE of second order. In Cartesian coordinates it was linear PDE of first order and we get closed form solution by characteristics method [2]. We see from (19), (28), (55), that this is possible only for Cartesian coordinates.

3. Possible applications of results of this paper are solution of Fokker - Planck equation by means of separation of variables method and formulation of boundary conditions on coordinate surfaces. We hope to give examples of these in following articles.

______________________________

**REFERENCES**

[1]  A. J. Mc Connel, Applications of tensor analysis. Dover Publications Inc., New York, 1957.

[2]  Igor A. Tanski. Fundamental solution of Fokker - Planck equation. arXiv:nlin.SI/0407007 v1 4 Jul 2004